# Identifying knowledge gaps in biodiversity data and their determinants at the regional level

Anaïs Guery[a], Didier Alard[a, b]

a. University of Bordeaux, US FAUNA, Pessac, France

b. University of Bordeaux, INRAE, UMR BIOGECO, Pessac, France

Acknowledgements

We thank the contributors of the Observatoire FAUNA database (organisms and individual contributors) whose data has been used for this project. We also thank the team of Observatoire FAUNA, especially Paul Fromage, for assisting in the data extraction and treatment.

Declaration of Interest

The authors of this manuscript declare no conflicts of interest.

Corresponding author address

Didier Alard : didier.alard@u-bordeaux.fr

University of Bordeaux, INRAE, UMR BIOGECO, Pessac, France

+33 (0)5 40 00 87 74

Data availability statement

Data used for this study include sensitive occurrence data, according to the Système d'Information de l'iNventaire du Patrimoine Naturel (SINP) guidelines. These data can not be released, unless there has been a motivated request by an identified individual. Therefore, data used for this study will only be made available on request.

**ABSTRACT**

Biodiversity open-access databases are valuable resources in the structuring and accessibility of species occurrence data. By compiling different data sources, they reveal the uneven spatial distribution of knowledge, with areas or taxonomic groups better prospected than others. Understanding the determinants of spatial and taxonomic knowledge gaps helps in informing the use of open-access data. Here, we identified knowledge gaps' determinants within a French regional biodiversity database, in the largest administrative region in France. Knowledge gaps were assessed using two metrics, completeness and ignorance scores, for 8 taxonomic groups covering five vertebrates and three invertebrates groups. The data was analyzed for the entire region, but also at the level of the three former sub-regions, to identify the potential drivers that may account for knowledge gaps' determinants. Our findings show that invertebrates were characterized by higher knowledge gaps than vertebrates. Overall, knowledge gaps are influenced by variables related to sites' accessibility rather than ecological appeal across both metrics. All groups shared similar determinants of gaps, except for the impact of agricultural pressure which is found to be more significant for invertebrates than vertebrates. Ultimately, our study emphasizes the impact of biodiversity governance, through local funding and regional political decisions, on knowledge distribution in open-access databases. We recommend limiting these biases by redirecting biodiversity funding towards under-sampled taxonomic groups and under-prospected areas. When not possible, users of data extracted from these databases should correct for spatial-sampling biases (SSP) using knowledge gaps' maps in order to get a more accurate understanding of species occurrence.

# 1. INTRODUCTION

As biodiversity declines worldwide, the need for knowledge about species distribution and biodiversity changes becomes increasingly pressing for monitoring, research and conservation policies (Butchart et al., 2010). The development of biodiversity databases in the last decades, has made biodiversity data more accessible and structured (Grassle, 2000 ; Reichman et al., 2011). This can be illustrated with the expansion of the Global Biodiversity Information Facility (GBIF), a world-wide open-access database created in 2001 and compiling over 3,000,000,000 species' occurrences in 2025 (GBIF, 2025). Such databases represent a valuable resource for research programs as they compile and standardize data which can be used to assess ecological dynamics such as species' distribution and population trends (Ball-Damerow et al., 2019 ; Marcer et al., 2022 ; Ribeiro et al., 2019). Ultimately, they support actions within the framework of public environmental policies.

However, open-access databases and their reliability for scientific research, especially regarding occurrence data, have been criticized for their inaccuracies and biases making them, in some cases, unfit to assess trends in species populations (Gaume & Desquilbet, 2024 ; Roel van Klink et al., 2020). Datasets have to be thoroughly examined to detect inaccuracies such as species' misidentification or discrepancies in projection systems (Colli-Silva et al., 2020 ; Freitas et al., 2020). Reviews point the uneven distribution of data due to spatial sampling bias and uncoordinated prospecting effort, leading to misinterpretation of species' distribution or population trends at global and local scales (Beck et al., 2014 ; Grattarola et al., 2020 ; Rocha-Ortega et al., 2021; Yang et al., 2013). Open-access databases are also shaped by taxonomic biases, with vertebrates' groups being better sampled than other taxa (Hugues et al., 2021 ; Garcia-Rosello et al., 2025 ; Troudet, Grandcolas, Blin et al., 2017). Biases in online databases impact potential data reutilization for scientific programs as they add several steps of data cleaning and treatment (Hill, 2012 ; Maldonado et al., 2015 ; Panter et al., 2020 ; Zizka et al., 2020). For instance, spatial sampling bias has to be corrected for species distribution models (SDMs) when using opportunistic data to improve reliability of models (Baker et al., 2024 ; Inman et al., 2021 ; Syfert et al., 2013). Thus, concerns have been raised about the use of occurrence data as low prospecting of certain areas or taxonomic groups causes spatial-sampling bias (Chesshire et al., 2023 ; Garcia-Rosello et al., 2023).

Two important challenges for the improvement of databases are to identify under-prospected areas and taxa and to understand the determinants of these gaps. Identifying spatial gaps may be performed via estimators of completeness and ignorance metric (Sousa-Baena et al., 2013 ; Girardello et al., 2019 ; Stropp et al., 2016 ; Ruete, 2015). Determinants of spatial knowledge gaps can be divided into two groups : variables related to sites' accessibility and variables related to the ecological appeal of sites. Sites' accessibility comprise density of road networks, distance to universities, degree of urbanization, etc (Correia et al., 2019 ; Mair & Ruete, 2016 ; Oliveira et al., 2016), while sites' ecological appeal include the distance to protected areas, the anticipated richness of habitats or attendance pressure (Meyer et al, 2015 ; Sanchez-Fernandez et al., 2008). However, studies usually focus on induvial species or taxa, although there is no evidence that these knowledge gaps are spatially congruent between taxonomic groups, particularly because the ecological attractiveness of sites depends on the species itself (Prendergast & Eversham, 1997). Indeed only a few studies have compared knowledge gaps' determinants across taxa while remaining confined to a given phylum or class (Sanchez-Fernandez et al., 2022 ; Šmíd, 2022).

Moreover, studies on knowledge gaps in biodiversity are mostly carried out on large spatial scales, at least at a national, continental or even global level. We believe that they can therefore only consider factors that are relevant at these scales and overlook more specific determinants that account for regional, cultural or geographical features affecting data collection. For example, in France, the open-access biodiversity program called the National Inventory of Natural Heritage (in French, SINP) was created in 2007 to improve the accessibility and centralization of French public biodiversity data (INPN, 2025 ; Poncet,

2013). It mostly relies on regional platforms, handling their own databases and transferring regionally collected data to the national database (Robert, 2021). The regional scale thus appears to be the preferred level for data collection, for the management of observers networks and is directly linked to regional nature conservation policies. The FAUNA Observatory collects and manages SINP data for the French administrative region Nouvelle-Aquitaine. In 2025, it compiles over 11,100,000 occurrence data, collected across 16,520 species (Observatoire FAUNA, 2025).

In this paper, we assess knowledge gaps in the FAUNA database across 8 taxonomic groups, including common vertebrates and insect groups. We then aim at identifying the underlying determinants of these gaps. Explicative patterns are analyzed through three prisms : (1) taxonomic, as we assume spatial incongruence of knowledge gaps across taxonomic groups, (2) bio-geographic, as we study the respective weights of the ecological attractiveness of sites or their accessibility in determining knowledge gaps, and (3) socio-administrative, because we aim at understanding the gaps in the regional database through the particular role of the 3 former regions which historically structured data collection, influenced the involvement of natural societies, defined taxonomic and spatial priorities on their respective territories, until their merger in 2015 to form the Nouvelle-Aquitaine region.

We use two different metrics of knowledge gaps to account for methodological issues. Ultimately, this study offers a comprehensive view of spatial sampling bias of the region, that can be used to inform further prospecting efforts and political decisions.

## 2. MATERIALS AND METHODS

*2.1. Study area*

The Nouvelle-Aquitaine region is a French administrative region, which was formed in 2015 by the political merging of three smaller regions : Aquitaine, Poitou-Charentes and Limousin (. 1). It stretches over 400 kilometers north of the Spanish border and counts 12 administrative sub-entities, called departments. In total, the study area covers 84,000 km² .

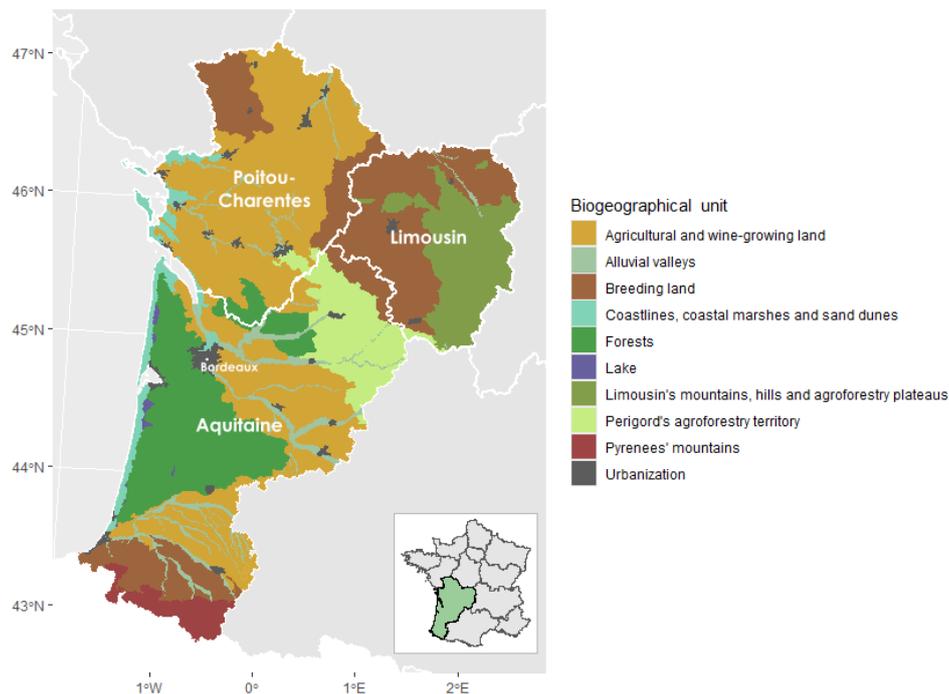

**Fig. 1** Map of the Nouvelle-Aquitaine region with its former constituent regions and the main biogeographical units.

The region's climate is temperate, tending towards oceanic. There is a vast variety of natural habitats, among which the Pyrenees' mountain range, coastlines mostly represent by sand dunes, forests, meadows and wetlands. However, it is semi-natural habitats, including agricultural and breeding land, that cover most of the region's surface. Despite being the largest French administrative region, Nouvelle-Aquitaine' is characterized by a rather low and heterogenous population density, generally higher in the municipalities located along the Atlantic coast and around large cities.

There are contrasts between the former regions. The former Limousin is mainly characterized by medium mountains with few large cities. The former Poitou-Charentes is essentially agricultural and coastal. Finally, landscapes are more diverse in the former Aquitaine, from mountains to the coast.

*2.2. Data selection*

Occurrence records were extracted from the FAUNA database on June 20$^{th}$ 2025 for 8 taxonomic groups : amphibians, bats, terrestrial birds, terrestrial mammals, reptiles, diurnal butterflies, odonates (dragonflies and damselflies), and orthopterans (grasshoppers, locusts and crickets) (FAUNA, 2025).

Due to high variations in data quality and precision, it was filtered using FME Workbench 2024.2.3 (FME Software, 2025) to keep only the records matching the following criterions :

- Temporal : Only records collected since 01/01/2010 are used. Records with a date span exceeding 1 month are removed.
- Spatial : Only the data entirely located in the Nouvelle-Aquitaine region are kept. Records located by a polygon with a surface area superior to 1 km$^2$ are removed, as for the records located by a linestring with a length superior to 1 kilometer.
- Taxonomic : Only records determined at the species level or below are used. Records at the infra-specific level are associated with the corresponding species.
- Reliability : Based on the FAUNA automatic validation system, the records with a reliability level labelled as 'doubtful' or 'very doubtful' are removed.

After filtering, the dataset includes 2,420,654 occurrence records.

Using a 10 x 10 kilometer grid of 856 cells, we associated each occurrence record with the corresponding cell. Records intersecting several cells were associated with the cell intersecting the highest proportion of the polygon's surface. To avoid any border effect, we removed all grid cells intersecting less than 40% of the Nouvelle-Aquitaine region.

*2.3. Spatial knowledge gaps assessment*

Two metrics were used to assess knowledge gaps per grid cell for each taxonomic group in the region: completeness, based on observed species, and ignorance index, based on the prospecting effort. They were carried out using R version 4.4.3 and integrated into a FME Workbench version 2024.2.3 workspace (R Core Team, 2025). Multitaxa scores were calculated for each metric by averaging all taxonomic groups' scores per grid cell.

### 2.3.1. Knowledge gaps based on the completeness metric

We calculated completeness for each grid cell using the Chao2 incidence-based index (Chao & Jost, 2012 ; Chao et al., 2020 ; Sousa-Baena et al., 2014 ; Troia & McMamanay, 2016). This index relies on species considered "rare" to estimate the number of undetected species ($S_{und}$) for each grid cell. For the purpose of this study, years were used as a proxy for samples. The unbiased following formula was used (Chao & Colwell, 2017 ; Gotelli & Colwell, 2011 ; OBIS, 2025) :

$$S_{und} = \left(\frac{m-1}{m}\right) \frac{q_1(q_1-1)}{2(q_2+1)}$$

where $S_{und}$: number of undetected species on the cell;

$m$ : total number of samples (15 years) ;

$q_1$ : number of species observed on one sampling unit (one year);

$q_2$ : number of species observed on exactly two sampling units (two years).

The number of undetected species is then used to estimate the expected number of species ($S_{exp}$) for each grid cell:

$$S_{exp} = S_{obs} + S_{und}$$

Where $S_{exp}$: number of species expected on the grid cell;

$S_{obs}$ : number of species observed on the grid cell.

The Chao2 index is not reliable for grid cells with few occurrence records. The number of singletons and doubletons is likely to be particularly high due to the lack of data, leading to an overestimation of the number of undetected species. Therefore, the number of expected species is not calculated for cells with less than 10 occurrence records.

The number of singleton and doubleton is highly correlated with sampling effort, which can cause high variations of the number of expected species within the same biogeographical unit. To be able to compare knowledge gaps within biogeographical units, the average number of expected species was calculated for each biogeographical area (Fig. 1). For this purpose, the majority biogeographical unit was calculated and associated with each 10x10 km cell. The average number of expected species is established using the following formula:

$$m_{déf} = \frac{\sum S_{att.i}}{N_i}$$

where $m_{def}$ : average expected species for the biogeographical unit;

$S_{att.i}$: number of species expected for the grid cell $i$ of the corresponding biogeographical unit;

$N_i$: total number of grid cells belonging to the corresponding biogeographical unit.

Each cell is then assigned the same average number of expected species within the same biogeographical unit. Cells with less than 10 observations are also assigned the average number of expected species of the corresponding biogeographical unit.

Finally, knowledge gaps based on completeness are calculated using the following formula:

$$G_{com} = \frac{S_{obs}}{m_{unep}}$$

where $G_{com}$ : knowledge gaps based on the completeness metric.

### 2.3.2. Knowledge gaps based on the ignorance index metric

To calculate knowledge gaps using the prospecting effort, we used the half-ignorance index algorithm (Ruete, 2015). This index was developed to estimate an ignorance score from occurrence data extracted from public databases such as GBIF or SINP. It is based on the assumption that species from the same taxonomic groups share similar biases, since observers are generally specialized on taxonomic groups rather than on particular species. Based on the number of observations recorded for each cell, it can be used to identify poorly prospected cells.

We first established the ratio between the number of species observed and the number of observations (species observation index) for each grid cell (Ruete, 2015):

$$S_i = N_i / R_i$$

Where $S_i$: species observation index for grid cell *i*;

$N_i$: number of occurrence records on the grid cell;

$R_i$: number of species recorded on the cell.

To assess the half-ignorance score, a threshold number of observations $O_{0.5}$ is used. It corresponds to the number of observations sufficient to assume that the absence of observations for a species is 50% chance due to its real absence on that grid cell and 50% chance of being due to the species being undetected (Ruete, 2015). As an asymptotic curve, the higher $O_{0.5}$, the slower ignorance scores will reach 0 (Ruete, 2015).

Here, 3 values of $O_{0.05}$ were tested : 1, 5 and 10. Little variation was observed in the results depending on the $O_{0.05}$ values so the intermediate value of 5 was used.

Once the index of species observations and the number of reference observations have been defined, the following formula is used to calculate the final half-ignorance score (Ruete, 2015):

$$G_{ign} = O_{0.5} / (N_i + O_{0.5})$$

With $G_{ign}$ : knowledge gap based on the ignorance index metric for the grid cell *i*.

Ignorance scores use a reversed gradient compared to completeness : knowledge gaps are higher when close to 1 with the ignorance score whereas they are higher when close to 0 using the completeness metric. In order to compare both metrics, the inverse of the ignorance score was calculated using :

$$G_{inv} = 1 - G_{ign}$$

For all the following analyses, the inverse of the ignorance score was used as a knowledge score. Thus, as completeness and knowledge scores increase and reach 1, knowledge gaps become weaker.

*2.4. Potential determinants explaining knowledge gaps*

In order to identify the variables that could drive spatial or taxonomic variations of knowledge gaps, eight potential determinants were selected, based on the availability of datasets for the study area, accounting for naturalness levels, biodiversity scores, proportion of protected areas, average road density, urbanization pressure, agriculture pressure, and attendance pressure. Each cell was attributed a value for each of the variables described above using ArcGIS Pro 3.2.0 (ESRI, 2024).

Naturalness has recently been mapped on the national territory using a 20-meters resolution and it is based on landscape proxies designed as three elementary components : biophysical integrity, spontaneity of process and spatio-temporal continuity (Guetté et al., 2021). The three rasters were downloaded from the UICN website. The dataset aggregating the three components into the potential naturalness was used. It was clipped to match the extent of the study area. Using the ArcGIS "Zonal statistics as Table" tool, the maximum value of naturalness for each 10km x 10km grid cell was extracted. We did not use the mean values of naturalness as the large cell size compared to the fine-grained resolution of the metric tends to smooth the data, resulting in little variations between cells. We assume that small habitat patches, which can be of interest to observers, can be differentiated using this landscape analysis method.

Biodiversity scores come from the Nouvelle-Aquitaine's hotspots project using species' occurrence data, species' distribution modelling and spatial priorization to identify the areas of high conservation priority at a 1 kilometer resolution (Collectif, 2021). For this study, the results of the Additive Benefit Function (ABF) of the Zonation method (Moilanen, 2007) were used as they highlight cells in species rich areas. Using the ArcGIS "Zonal Statistics as Table" tool, the mean value of cumulative biodiversity was extracted for each 10 x 10 km cell. Intuitively, biodiversity hotspots relate more to ecological appeal than sites' accessibility, especially because cumulative richness is calculated from modeled distributions. However, the methodology of the Nouvelle-Aquitaine hotspots' study is strongly influenced by the visiting of sites and tends to highlight the most visited sites. It is at the intersection between attendance pressure and naturalness.

Average road density is used as a proxy for site accessibility. It was calculated using the "VOIE NOMMEE" shapefile of the national BD TOPO dataset (Institut Géographique National, 2025). The file was clipped to the extent of Nouvelle-Aquitaine's boundaries and the "Line density" tool was parameterized with a Search Radius of 5,000 meters and an output cell size of 1 kilometer. The "Zonal Statistics as Table" tool was then used to get the average road density for each cell.

Protected areas were identified as a determinant of knowledge gaps as observers are more likely to visit areas with a higher ecological values. Moreover, protected areas may act as site(cell) appeal as well as local actors of biodiversity governance, assuming that they can lead particular biodiversity surveys on their territories and collect funds for or initiate citizen science studies. Datasets of different ecological protection types were merged into one : national natural parks, ('Parcs nationaux'), regional natural parks ('Parcs Régionaux'), national natural reserves ('réserves naturelles nationales'), regional natural reserves ('reserves naturelles regionales') and 'sites of special fauna or flora interest ('ZNIEFF') (INPN, 2025 ; Magellium, 2025 ; Observatoire FAUNA, 2025 ; Parcs Nationaux de France, 2025). The tool "Tabulate Intersection" was then used to calculate the surface's percentage of each grid cell covered by protected areas.

Anthropogenic pressures with a significant impact on biodiversity were mapped in 2021 as part of a French national research program by the Museum unit PatriNat (Cherrier et al., 2021). Summary datasets aggregate pressures into three categories at the 10 kilometers resolution : agriculture, urbanization and attendance. The rasters were downloaded from the PatriNat website and clipped to the extent of the Nouvelle-Aquitaine region. As the same 10 km x 10 km grid is used by the two studies, no further processing was performed. The pressure values for each of the three category was joined to each cells based on a common identifier.

*2.5. Statistical analyses*

In order to identify potential congruences in knowledge gaps between the eight taxonomic groups, two PCAs were performed, one for each metric on a [856 cells x 8 scores] data matrix. For both metrics, completeness and knowledge, cell scores are at the 10x10km grid resolution. We examined whether the ex-regional level show different patterns due to contrasting survey methods and governance. To split grid cells according to the three ex-regions, the ArGIS "Tabulate Intersection" tool was used between the grid cell dataset and the shapefile of ex-regions contours. Each grid cell was joined to the ex-region intersecting most of its surface. PCA were then conducted in R 2024.12.1 on the datasets using the *FactoMineR* package (Le et al., 2008 ; R Core Team, 2025).

We explored the spatial potential determinants of knowledge gaps by the mean of a PCA on a [856 cells x 7 determinants]. This analysis allows us to determine whether the heterogeneity of land covers and geographic features, especially within and between ex regions, translates into a diversity of determinants combinations.

To study the relationships between potential determinants and groups' knowledge gaps, redundancy analyses (RDA) were conducted for vertebrates, invertebrates and across all taxonomic groups at the regional level. The rda() function of the *vegan* package was used (Oksanen et al., 2025).

Taxa were merged into two broader categories based on similarities in knowledge gaps identified through the PCA: vertebrates excluding bats (amphibians, reptiles, terrestrial birds and non-flying mammals) and invertebrates (odonates, orthopterans, diurnal butterflies). For each category, knowledge gaps scores of the included taxonomic groups were averaged for each grid cell using the "Statistics Calculator" tool in FME Workbench (FME Software, 2025). Mean scores were calculated using completeness for one dataset and knowledge index for the other.

To study relationships between variables and knowledge gaps for each individual taxonomic groups, linear regressions were conducted using the *stats* package (R Core Team, 2025). They were performed for the two knowledge gaps metrics for each ex-subregion and the whole area for individual taxonomic groups as well as for vertebrates and invertebrates.

All analyses were carried out using R 2024.12.1 (R Core Team, 2025).

## 3. RESULTS

*3.1. Knowledge gaps' spatial and taxonomic distribution*

Both metrics display similar spatial distribution (Fig. 2), with lowest completeness and knowledge scores located in the north-west of the region, that is to say within the territory of the former Poitou-Charentes region. Within each metric, scores are similarly distributed between vertebrates and invertebrates. Invertebrates have overall higher knowledge gaps compared to vertebrates (Fig. 2). Scores for all taxa is more evenly distributed and reduces contrasts and distribution heterogeneity.

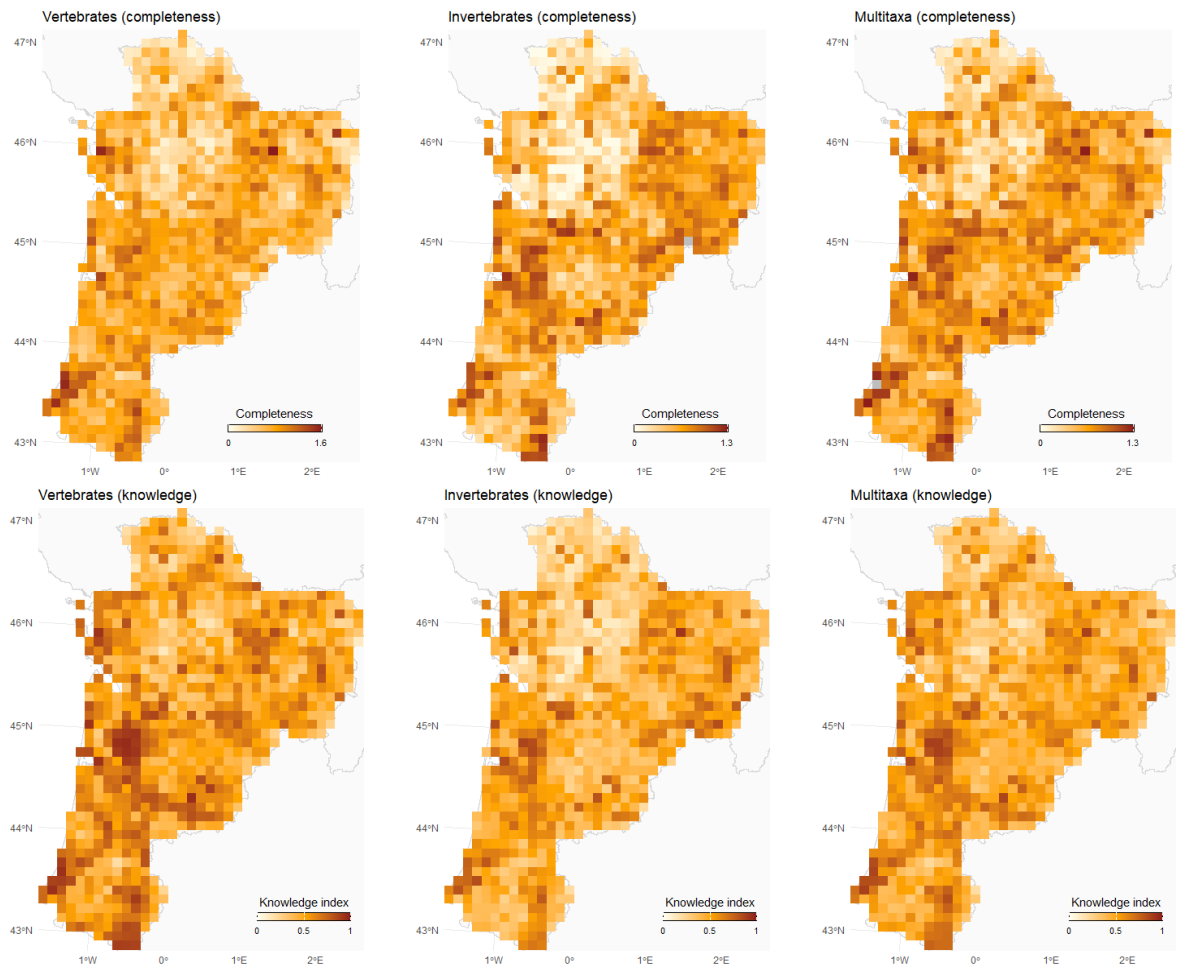

**Fig. 2** Spatial distribution of knowledge and completeness scores for vertebrates, invertebrates and all taxa using a 10 km x 10 km grid cell of Nouvelle-Aquitaine. Completeness scores are represented can be higher than 1 as observed species can exceed the number of expected species. This gradient allows to depict areas where that is the case, as well as the least-sampled areas. On the other hand, knowledge scores are constrained on a scale from 0 to 1.

Knowledge gaps' patterns exhibit heterogeneity within and between ex-regions, suggesting a difference in the origin of knowledge gaps. In the north-west, the ex-region Poitou-Charentes is identified with overall lower knowledge and completeness scores across all groups and metrics, especially for invertebrates (Fig. 2 / Supplementary material Figure S1 and Figure S2). Cells located at the eastern and northern border of the region have overall higher knowledge gaps. Taxonomic groups are characterized by distinct distributional patterns in the region. For instance, birds are being overall better prospected with high completeness scores near the coastline which is not the case of other taxa (Supplementary materials Figure S1 and Figure S2).

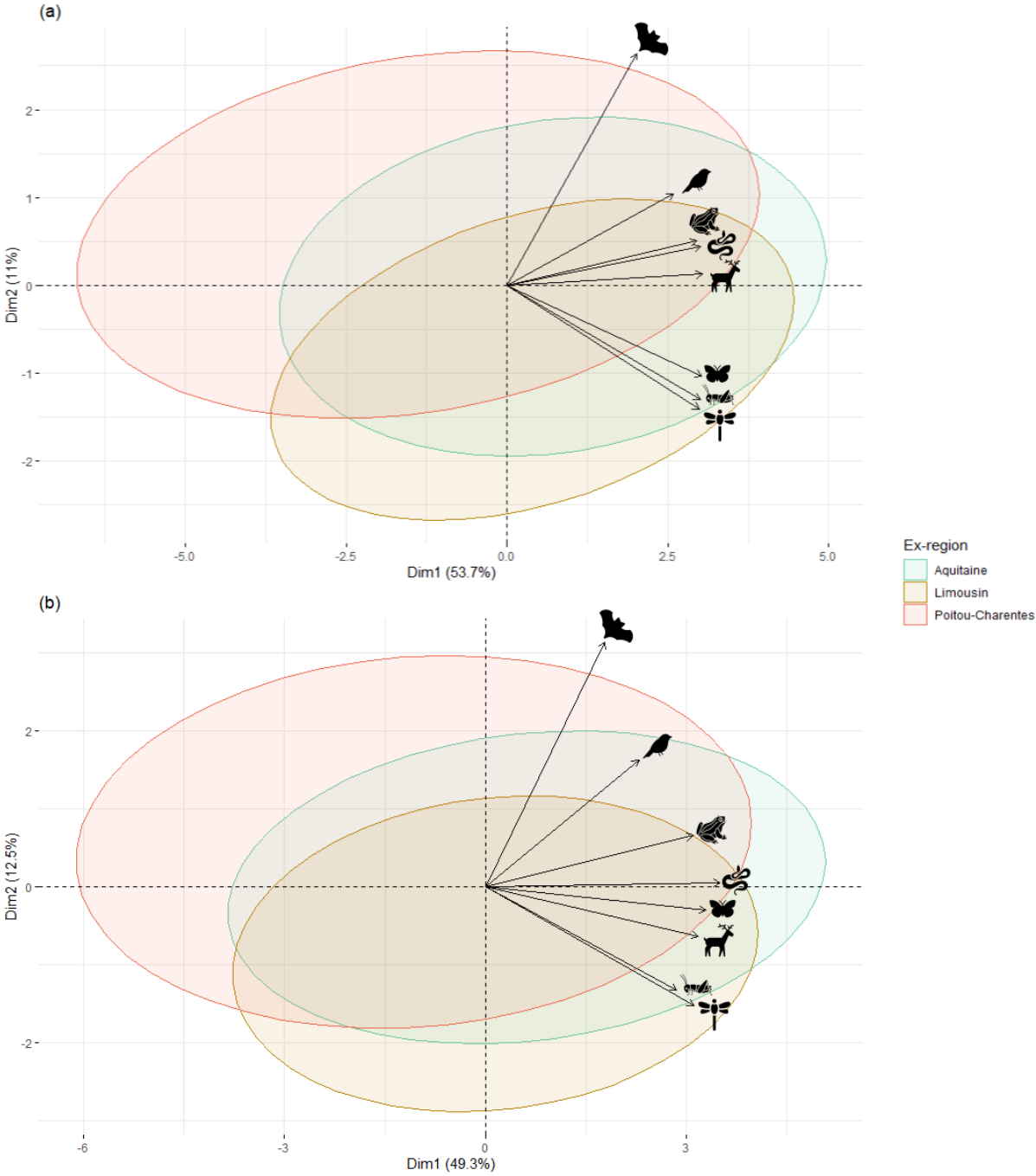

**Fig. 3** Principal component analyses (PCA) biplots on a matrix of 856 grid cells and 8 completeness (Fig. 3a) and knowledge scores (Fig. 3b) per taxonomic groups, from highest to lowest in the second dimension (a) : bats, terrestrial birds, amphibians, reptiles, terrestrial mammals, diurnal butterflies, orthopterans, odonates. The directions and lengths of the vector arrows show the loadings of each taxa's score to the principal components. Ellipses account for grid cells' distribution within each ex-sub region using centroids and a 90% confidence interval.

PCAs performed on knowledge and completeness scores for each taxonomic group give consistent results (Fig. 3). Considering the first PCA based on completeness (Fig. 3a), the first dimension explain 53.7% of the total variance while dimension 2 explains 11%.

Two clusters of taxonomic groups can be identified from the analysis on completeness scores : a first group of vertebrates with birds, amphibians, reptiles, mammals on one hand and a second group with invertebrates including odonates, orthopterans and diurnal butterflies on the other hand. Invertebrates' scores are orthogonal to bats'. Vertebrates are positively correlated with dimension 1 only while invertebrates are positively correlated with the first dimension but negatively with the second dimension.

As a whole, the further a grid cell is positioned towards the position part of the first axis, the better the knowledge. Conversely, positions towards the negative part of this axis indicate a high level of knowledge gaps at the cell level. The nature of these gaps, that is, the taxa considered to have incomplete information in the cell, influences the position of cells on the second axis of the PCA. Therefore, these gaps primarily concerns bats at the top, the positive part of axis 2, and invertebrates (insects) in the negative part of this second axis.

Regarding the ellipses visualizing the distribution of cells of the former regions, their average position on axis 1 describes the average level of completeness, while their position on the second axis visualizes the nature of the taxa responsible for the knowledge gaps. The size of the ellipses, in turn, describes the variability of these levels and the nature of the gaps within the cells of the former regions. Therefore, the distribution of ellipses shows a disparity of gaps between ex-regions, with Aquitaine showing higher completeness on average. The scores are lower on average for Poitou-Charentes, which displays a greater heterogeneity of situations.

Quasi similar patterns can be observed in the second PCA based on knowledge scores (Fig. 3b). Its first dimension explains 49.3% of the total variance while the second dimension explains 12.5%. It differs slightly from the PCA on completeness as butterflies are more linked with the dimension 1 and are therefore closer to vertebrates' knowledge distribution. Ex-regions' PCA positioning are similar for knowledge and completeness score (Fig. 3).

*3.2. Identifying knowledge gaps' determinants*

The two first dimensions of PCA on knowledge gaps determinants explain about 58% of the variation (Fig. 4). Most of the variation is explained by the first dimension, close to 37%. Naturalness and protected areas are strongly correlated with dimension 1, and are utterly opposed primarily to agricultural pressure, which is correlated with the negative part of the first dimension. Biodiversity hotspots are mostly correlated to gradients of naturalness and protected areas, they appear to be at the intersection between attendance pressure and naturalness. As a whole, determinants related to habitat quality (naturalness, protected areas and hotspots) are logically opposed to variables related to human pressure (agriculture and urbanization).

Ex-subregion Aquitaine displays more dispersal of its cells in the determinants' space, in accordance with its high habitat heterogeneity, compared to a more homogeneous ex-region such as Poitou-Charentes. Apart from that, there are similar mean positions of grid cells in the determinants' space for the three different ex regions insofar as the three ellipses are all centered on the variable space, suggesting that although contrasts are higher in some ex-regions, they are of same nature within each ex-region.

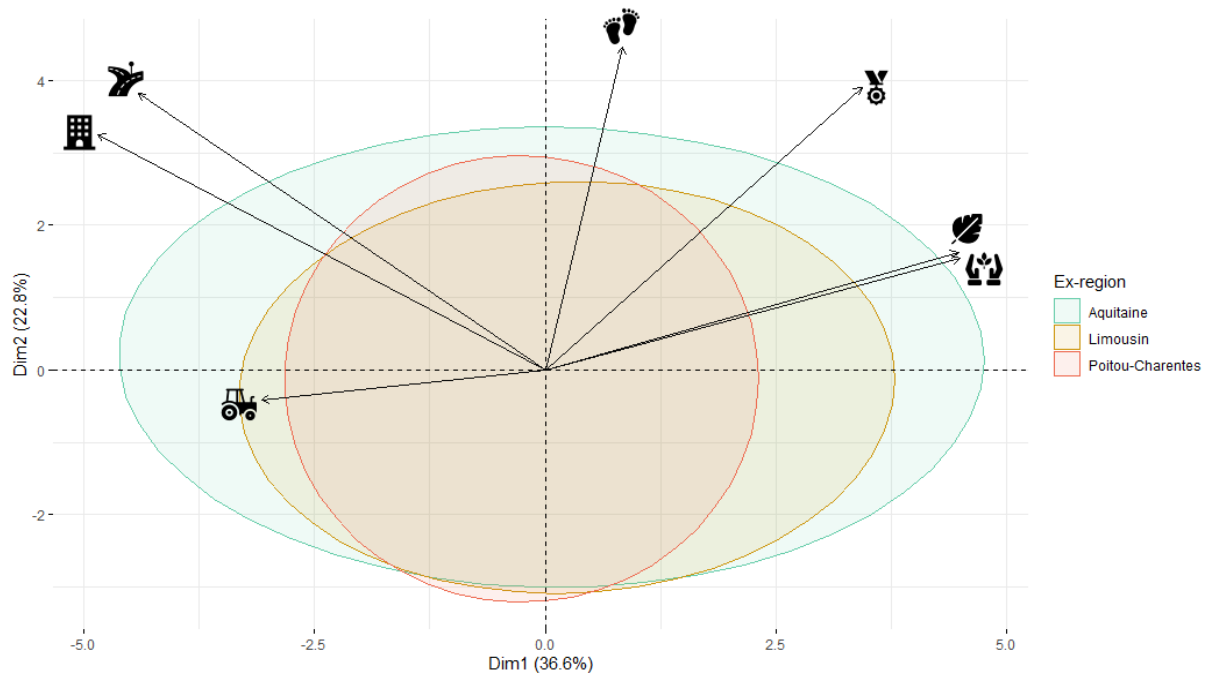

**Fig. 4** Principal component analysis (PCA) biplot on a matrix of 856 grid cells and 7 determinants of knowledge gaps in Nouvelle-Aquitaine, from left to right in the first dimension : urbanization pressure (building), road density (roads), agricultural pressure (tractor), attendance pressure (footprints), hotspots (medal), naturalness (leaf) and surface of protected areas (hands). The directions and lengths of the vector arrows show the loadings of each determinants to the principal components. Ellipses account for grid cells' distribution within each ex-sub region using centroids and a 90% confidence interval.

Knowledge gaps' determinants exhibit different patterns within vertebrates and invertebrates groups (Fig. 5, Fig. 6), confirming different observation biases between the two subphylum. Magnitude of the relationships and statistical significances of variables differ between ex-regions within each taxonomic group (Fig. 6). Adjusted $R^2$ are low, the selected variables explaining only part of the variations in knowledge gaps.

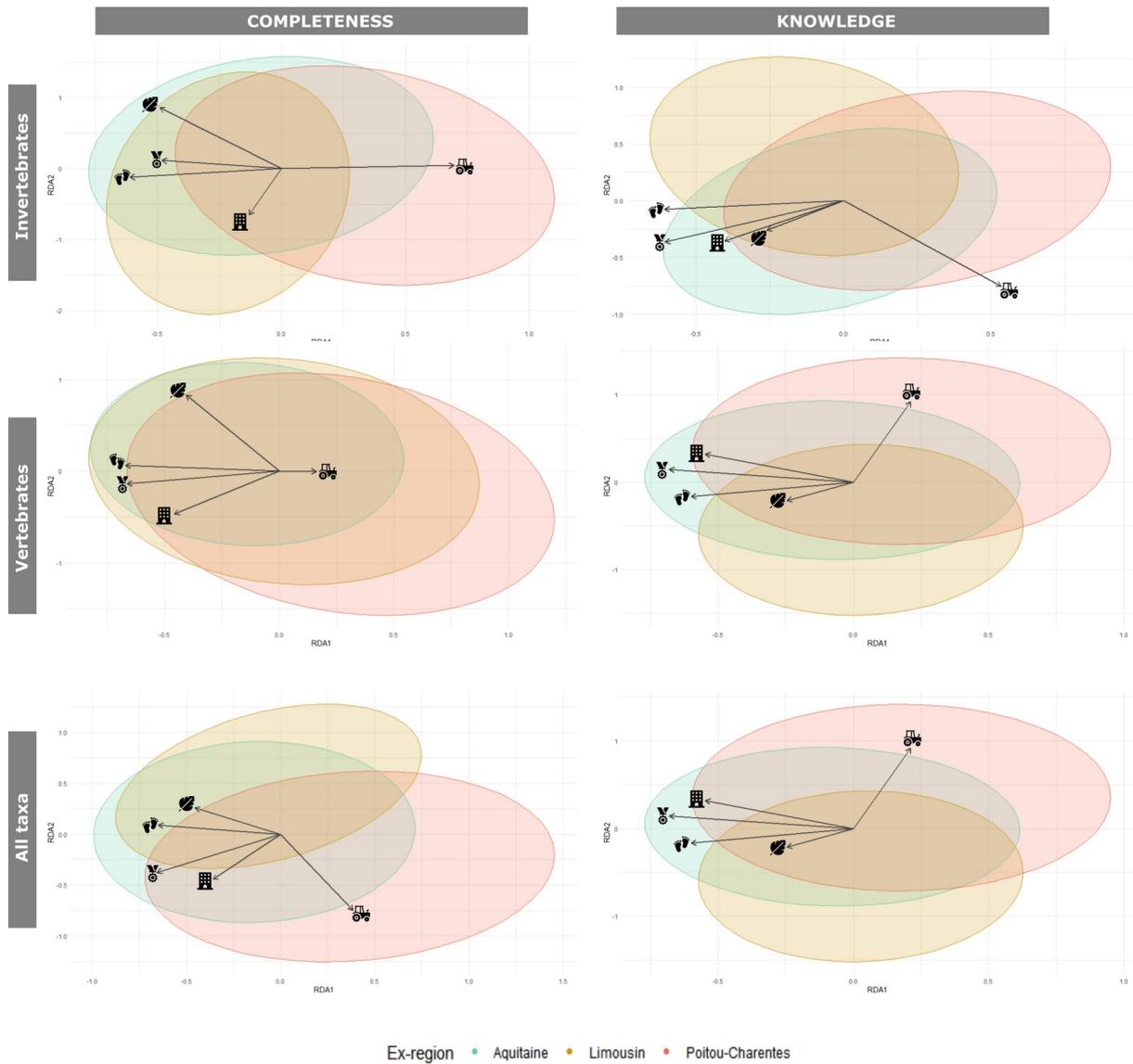

**Fig. 5** Redundancy analyses (RDA) showing the contribution of knowledge gaps' determinants to completeness and knowledge scores for vertebrates, invertebrates and all taxa. The directions and lengths of the vector arrows show the strengths of correlations between knowledge gaps' scores and determinants.

Knowledge gaps are mostly negatively explained by hotspots density and urbanization pressure. Overall, biodiversity hotspots have the stronger significant contribution to knowledge gaps, comprised between 0.3 and 0.6, meaning that knowledge gaps are lower in biodiversity hotspots (Fig. 6). Urbanization pressure is also a significant determinant for most of the taxonomic groups, with knowledge gaps declining in more urbanized areas. Its contribution is generally moderate, comprised between 0.02 and 0.07. Attendance pressure is found across all groups and ex-regions, having a positive contribution to knowledge gaps, despite its weak magnitude.

Other variables are overall less significant determinants of completeness and knowledge. When significant, agricultural pressure is the only variable on the positive end of RDA 1, having a negative contribution to knowledge and completeness scores, which are decreasing when agricultural pressure increases (Fig. 5). Indeed, having a positive influence on the first dimension, agricultural pressure is orthogonal to other determinants, which all are on the negative side of RDA 1 (Fig. 5).There is an exception with amphibians' knowledge score in Limousin, where agricultural pressure has a positive contribution (Fig. 6). Other variables display most of their variation in the RDA 2, such as naturalness and urbanization pressure, which positively influence completeness and knowledge scores (Fig. 5).

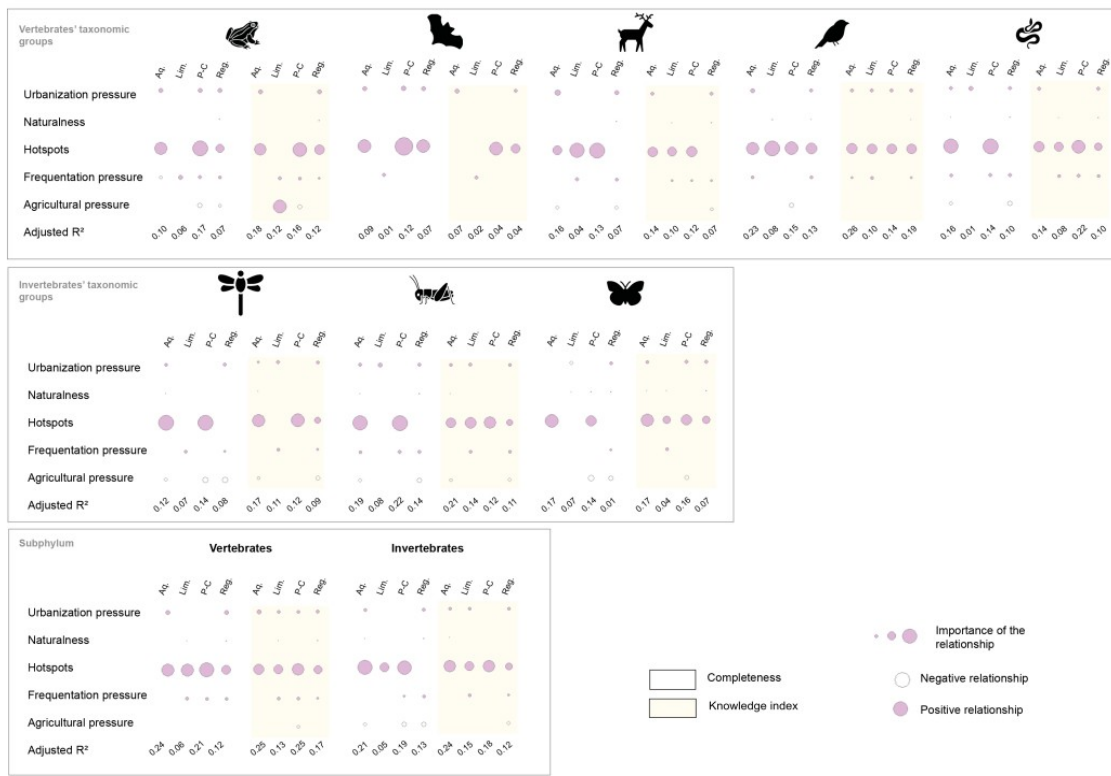

**Fig. 6** Results of linear regression models between knowledge gaps' scores and determinants for each taxonomic group and subphylum. Only variables with p-value <0.05 are represented here. The size of the circle represents the strength of the relationship. Pink circles indicate positive relationships while white circles display negative relationships.

Knowledge gaps' determinants differ between taxonomic groups. For vertebrates, there are similarities between reptiles and amphibians, which are mostly driven by hotspots and attendance pressure (Fig. 6). Birds and mammals exhibit more variations as they are mostly influenced by hotspots and urbanization pressure. Invertebrates' scores are driven by hotspots and urbanization pressure but also by agricultural pressure, which has a stronger significant contribution than for vertebrates (Fig 5, Fig. 6).

## 4. DISCUSSION

From the assessment of knowledge gaps, based on two different measures of sampling effort, in an open-access online regional database, we were able to identify underlying patterns using three perspectives : geographic, with the spatial distribution of gaps and their determinants ; taxonomic, looking for differences between taxonomic groups ; and socio-administrative, with the weight of ex-regions and biodiversity governance. Important geographic and taxonomic disparities were found, with consistent findings across both metrics. Spatial distribution of knowledge gaps is heterogeneous among ex-regions and taxonomic groups, resulting in several poorly sampled areas and taxa. There tends to be more similarities when comparing groups belonging to the same subphylum, vertebrates or invertebrates. Across all studied taxonomic groups, knowledge gaps in Nouvelle-Aquitaine are more influenced by sites' accessibility than sites' ecological appeal. Biodiversity hotspots and urbanization pressure display the strongest and most consistent influence on sampling effort. However, vertebrates and invertebrates' determinants exhibit differing patterns, especially when it comes to the relationship with agricultural pressure.

*4.1. Patterns of knowledge gaps : the impact of taxonomic bias*

Knowledge gaps' comparisons across taxonomic groups exhibit differing results based on the selected scale.

When grouping taxa as vertebrates and invertebrates, knowledge gaps' distribution display similar spatial patterns across both groups. Invertebrates have overall weaker knowledge and completeness scores than vertebrates, which are likely the result of taxonomic biases. Indeed, vertebrates' species are overall better sampled by professional inventories as well as citizen science project as they spark more interest (Petersen et al., 2021 ; Phaka et al., 2022). They are also deemed easier to sample than invertebrates, for which species might require specific sampling techniques and laboratory analyses to be identified (Doxon et al., 2011 ; Sittenthaler et al., 2023). It results in weaker knowledge gaps for invertebrates in open-access databases, which is confirmed by the results of scientific literature (Clark & May, 2002 ; Collen et al., 2008 ; Correia et al., 2019 ; Titley et al., 2017).

Nevertheless, the comparison of sampling bias across each taxonomic group demonstrates stronger variation in the magnitude of completeness and knowledge scores as well as knowledge gaps' distribution, confirming that some taxonomic groups are better sampled than others (Rocha-Ortega et al., 2021). For instance, birds are overall better prospected than bats and orthopterans with best sampling close to the Atlantic coastline while amphibians' higher scores are located near urbanized areas. There may be an influence of observers' site selection bias, being more likely to sample favorable habitats which are not necessarily located in the same areas across taxonomic groups (De Araujo & Ramos, 2021 ; Troia & McManamay, 2016 ; Mentges et al., 2020).

Overall, there are similar spatial sampling determinants across taxonomic groups, at the exception of agricultural pressure. Having a significant relationship for only a few vertebrates' taxonomic groups across a few ex-regions, it influences invertebrates' knowledge gaps across the three groups and for some ex-regions. Agricultural pressure mostly has a negative relationship with knowledge and completeness scores, which decrease when agricultural pressure increases. Invertebrates such as butterflies and dragonflies are mostly associated with open habitats, including meadows and fields. Agricultural areas being open habitats, they are sampled to investigate invertebrates' richness and diversity, suggesting that knowledge gaps would be weaker in agricultural areas (Dolny et al., 2021). But chemical treatments and human exploitation of these semi-natural habitats have also been shown to negatively influence butterflies and other invertebrates' populations (Rands & Sotherton, 1986 ; Sanchez-Bayo, 2021). Thus, it is likely that within accessible sites, naturalists prioritize habitats favorable to the taxonomic group (Andrade-Silva et al., 2022 ; Boakes et al., 2010 ; Martin et al., 2012). That would include open-habitats with little human impact for invertebrates and closed habitats, such as forests, for vertebrates.

Bats have exhibited different patterns than the other taxonomic groups, which singularity allows to illustrate variations of knowledge gaps' dynamics for the Nouvelle-Aquitaine region. Bats' sampling has been shown to be strongly influenced by spatial biases, causing heterogenous prospecting effort (Fisher-Phelps et al., 2017). Knowledge gaps' distribution does not display any ex-region structuring, suggesting that other biases are at play. While vertebrates' and invertebrates' observers might sample several taxonomic groups, bats are mostly sampled by chiropterologists, using species sampling techniques such as acoustic detection (O'Farrell & Gannon, 1999). Based on the habitat-selection bias, they mostly sample areas known to be favorable bats' habitats such as caves, buildings or trees, explaining why urbanization pressure has a slightly stronger influence on bats' knowledge gaps. Therefore, bats exhibit different knowledge gaps' patterns due to community-specific sampling methods.

*4.2. Spatial-sampling bias and knowledge gaps distribution*

Surprisingly, no significant variation has been observed in knowledge gaps' distribution or determinants among knowledge and completeness scores. As sampling effort is shaped by observers' biases, observers tend to focus more on rare and keystone species than on common species (Boakes et al., 2010 ; Garcillan et al., 2008 ; Gardiner et al., 2012 ; Minteer et al., 2014). It leads to areas displaying high numbers of observations for only a small subset of species, which would result in disparities between both metrics. In open-access databases, it causes rare species to have a disproportionately high number of observations compared to the size of the population. However, prospecting bias does not seem to impact the assessment of knowledge gaps as our findings follow the logic of species accumulation curves, with the number of observed species increasing with the number of samples (Ugland et al., 2003). It suggests that the number of observations and the number of species could then be used interchangeably as proxies of spatial sampling bias.

Across both metrics and all taxa, general patterns of knowledge gaps' regional distribution were identified. The north of the region, especially Poitou-Charentes, exhibits overall stronger knowledge gaps, while areas near the coastline and in the Pyrenees' mountain range are better prospected. Poitou-Charentes, being mainly agricultural land, is often associated with lower ecological stakes than mountainous and coastal habitats, where the numbers of endemic and vulnerable species are higher. It results in habitat-selection bias, as observers tend to visit areas where expected species richness, diversity and ecological stakes are high (Dennis & Thomas, 2000 ; Mentges et al., 2020).

When exploring knowledge gaps determinants at a finer scale, spatial sampling biases are found to be driven by sites' accessibility rather than sites' ecological appeal across all ex-regions and metrics. Urbanization pressure, hotspots and attendance pressure display the strongest and most consistent relationships with knowledge gaps, increasing as knowledge gaps increase. Even when other variables were used as proxies, accessibility has been shown to strongly influence spatial sampling (Andrade-Silva et al., 2022 ; Correia et al., 2019 ; Girardello et al., 2018 ; Mair & Ruete, 2016 ; Sobral-Souza et al., 2024). Areas close to cities, with a developed road network and higher human densities are overall better prospected. However, accessible areas are often correlated with anthropized areas, leading to better sampling in potentially ecologically poorer regions (Collen et al., 2008 ; Ronquillo et al., 2020). Ultimately, significant levels of knowledge are linked to the interest of the sites, likely to attract visitors, or to sites that are more visited because they are easily accessible and close to cities. Areas' richness being used to assess spatial and conservation priorities, said priorities could be over or underestimated due to this spatial sampling bias (Grand et al., 2007 ; Grattarola, Martinez-Lanfranco, Botto et al., 2020 ; Engemann et al., 2015). It is suggested that better-informed decisions could be taken by correcting species' distribution using assessments of spatial knowledge gaps.

Only parts of knowledge gaps' variation is explained by these determinants as adjusted $R^2$ values are relatively low across all taxonomic groups and ex-regions. Other studies exploring spatial sampling bias determinants have had similar results (Correia et al., 2019 ; Mair & Ruete, 2016). When using ignorance scores, it has been shown that the value of the ignorance's $O_{O.O5}$ parameter causes $R^2$ variation, which decreases when $O_{0.05}$ increases (Correia et al., 2019 ; Mair & Ruete, 2016). However, these are mostly small variations as $R^2$ values stay in a similar range, only differing in their absolute values.

*4.3. How biodiversity governance shapes regional knowledge gaps*

Part of the unexplained variance in the region could be explained by the political and historical structuring of the SINP program in Nouvelle-Aquitaine. Even though more and more data is collected by online public databases, ex-regions dynamics still strongly influence data collection by the FAUNA database (Touroult et al., 2020). Each ex-region had differently structured naturalist networks and functioning regarding biodiversity data before the merging. Some networks were already sharing their data with the SINP program while others were sharing with private databases or creating their own database to structure their data. The merging of the ex-regions did not lead to the cohesive structuration of a Nouvelle-Aquitaine database, which would require the formatting and transfer of millions of historical data. Instead, most naturalist networks are still committed to their initial method of data structuration which explains different practices regarding data transfer to the FAUNA database.

Naturalist structures of Poitou-Charentes transfer mostly data funded by public funds to the FAUNA database while Aquitaine's naturalists are also likely to transfer privately funded data. For instance, the average number of records per grid cell for terrestrial mammals is 288 records in Poitou-Charentes, 378 in Limousin and 464 in Aquitaine (FAUNA, 2025). Therefore, knowledge gaps' spatial distribution might be influenced by observers likelihood of sharing data on the regional SINP platform. These data sharing gaps across the Nouvelle-Aquitaine region are likely to be the strongest determinants of knowledge gaps variations across ex-regions. Sharing disparities are unaccounted for in this study as estimating recorders' propensity to transmit datasets is a difficult and controversial task.

Biodiversity sampling is not only shaped by interests as funding also plays a high role in deciding what species and regions should be prospected. Fauna-related projects are funded heterogeneously, focusing on specific areas and taxonomic groups, likely to lead to disparities across groups and territories (Martin-Lopez et al., 2009 ; Titley et al., 2017). In France and Nouvelle-Aquitaine, birds are well prospected, with historical participative programs such as STOC, which guarantee long-term datasets and an overall better spatial coverage of data (Vimont et al., 2025). For other taxonomic groups, especially invertebrates, surveys' funding is more likely to be temporally and spatially scattered, causing knowledge gaps at the regional and subregional scale. Within taxonomic groups, conservation funding has been shown to focus on charismatic species more than threatened species, as they are more likely to get funding and public attention (Guénard et al., 2025). Assuming that the same mechanisms motivate the funding of data acquisition programs, there is an urgent need for biodiversity governance and stakeholders to include species vulnerability and knowledge gaps assessments in the decisions for funding allocations.

At the spatial level, fundings also influence knowledge gaps variations : for most groups, cells located in the National Park of the Pyrenees display better sampling than other areas, due not only to the site's appeal but also to the internal funding of scientific programs. More generally, in the Nouvelle-Aquitaine region, national and regional parks are important contributors to biodiversity data, as illustrated by the spatial distribution of knowledge gaps. Indeed, French protected areas receive regular and consequent funding for biodiversity projects, especially by the French government. It causes strong disparities in knowledge gaps distribution between better sampled funded areas and areas receiving little to no funding.

Finally, more than funding, biodiversity sampling and preservation has been shown to be strongly correlated with political decisions (Le Velly et al., 2024 ; Magnusson et al., 2018). Protected areas boundaries are ultimately decided by stakeholders, strongly influencing the location of biodiversity conservation and surveys. Moreover, the temporal mismatch between short political terms and the need for long-term conservation policies impacts biodiversity programs and longevity (Steinberg, 2009). Therefore, knowledge gaps' determinants should further be explored in the light of socio-administrative drivers in order to improve biodiversity knowledge.

As demonstrated, knowledge gaps and their determinants are complex, intricated and multifactorial. To better explore and understand underlying dynamics, further investigations would be need. Temporal dynamics of data acquisition should be studied to identify further sampling biases. Indeed, data could be spread uniformly across the time period or clustered around 1 or 2 years, which would highly influence the understanding of species' distribution or ecology. Mapping knowledge gaps per year would allow to explore such variations. Similarly, species' phenology varies throughout the year, with many field surveys being conducted during species' periods of reproduction. Knowledge gaps would then be weaker during that time frame than during the rest of the year. Finally, at the intersection between taxonomic bias and temporal bias, it has been shown that naturalists tend to focus on specific life stages, sampling more adult individuals than exuviae for odonates for instance (Raebel et al., 2010). These different dynamics could have high impact on knowledge gaps' distribution and determinants. Improving our understanding of knowledge gaps' intricated patterns would allow better use of open-access data. Ultimately, understanding the key determinants of biodiversity knowledge gaps will require a better understanding of knowledge gaps in the data itself.

**STATEMENTS AND DECLARATIONS**

*Funding*

This work was supported by the Direction Régionale de l'Environnement, de l'Aménagement et du Logement Nouvelle-Aquitaine through funding of US FAUNA.

*Competing Interests*

The authors have no relevant financial or non-financial interests to disclose.

*Data availability*

Data used for this study include sensitive occurrence data, according to the Système d'Information de l'iNventaire du Patrimoine Naturel (SINP) guidelines. These data cannot be released, unless there has been a motivated request by an identified individual. Therefore, data used for this study will only be made available on request.

**FIGURE CAPTIONS**

**Fig. 1** Map of the Nouvelle-Aquitaine region with its former constituent regions and the main biogeographical units.

**Fig. 2** Spatial distribution of knowledge and completeness scores for vertebrates, invertebrates and all taxa using a 10 km x 10 km grid cell of Nouvelle-Aquitaine. Completeness scores are represented can be higher than 1 as observed species can exceed the number of expected species. This gradient allows to depict areas where that is the case, as well as the least-sampled areas. On the other hand, knowledge scores are constrained on a scale from 0 to 1.

**Fig. 3** Principal component analyses (PCA) biplots on a matrix of 856 grid cells and 8 completeness (Fig. 3a) and knowledge scores (Fig. 3b) per taxonomic groups, from highest to lowest in the second dimension (a) : bats, terrestrial birds, amphibians, reptiles, terrestrial mammals, diurnal butterflies, orthopterans, odonates. The directions and lengths of the vector arrows show the loadings of each taxa's score to the principal components. Ellipses account for grid cells' distribution within each ex-sub region using centroids and a 90% confidence interval.

**Fig. 4** Principal component analysis (PCA) biplot on a matrix of 856 grid cells and 7 determinants of knowledge gaps in Nouvelle-Aquitaine, from left to right in the first dimension : urbanization pressure (building), road density (roads), agricultural pressure (tractor), attendance pressure (footprints), hotspots (medal), naturalness (leaf) and surface of protected areas (hands). The directions and lengths of the vector arrows show the loadings

of each determinants to the principal components. Ellipses account for grid cells' distribution within each ex-sub region using centroids and a 90% confidence interval.

**Fig. 5** Redundancy analyses (RDA) showing the contribution of knowledge gaps' determinants to completeness and knowledge scores for vertebrates, invertebrates and all taxa. The directions and lengths of the vector arrows show the strengths of correlations between knowledge gaps' scores and determinants.

**Fig. 6** Results of linear regression models between knowledge gaps' scores and determinants for each taxonomic group and subphylum. Only variables with p-value <0.05 are represented here. The size of the circle represents the strength of the relationship. Pink circles indicate positive relationships while white circles display negative relationships.

**Fig. S1** Distribution of knowledge gaps for each taxonomic group using the completeness metric.

**Fig. S2** Distribution of knowledge gaps for each taxonomic group using the knowledge metric.

**SUPPLEMENTARY MATERIAL**

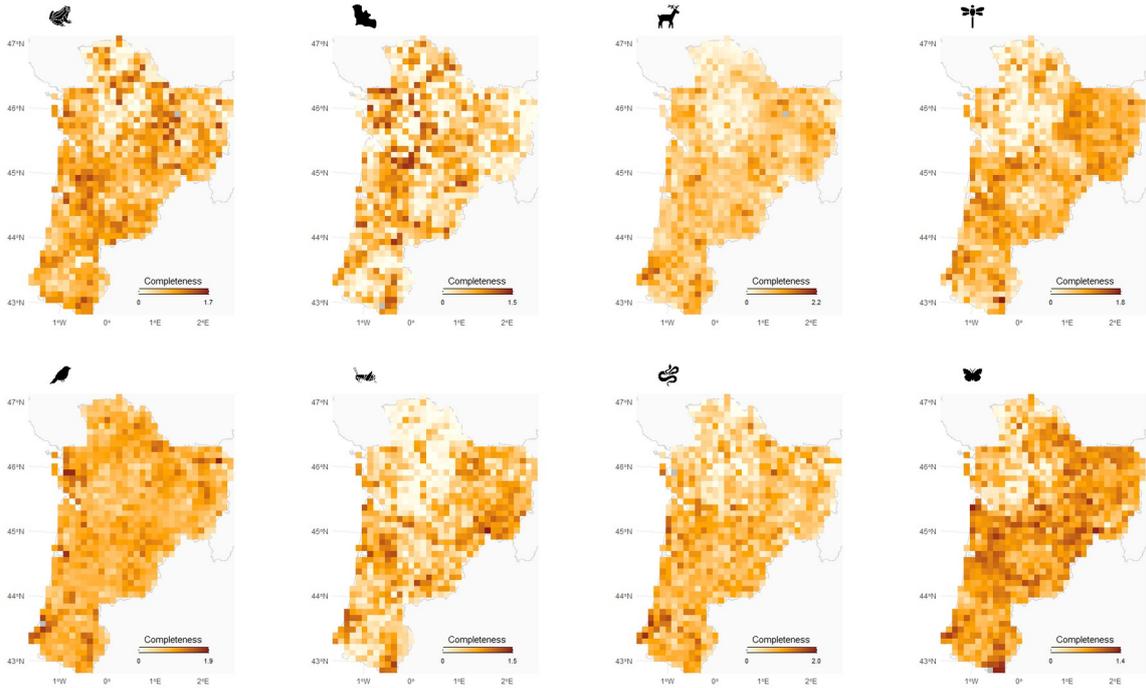

**Fig. S1** Distribution of knowledge gaps for each taxonomic group using the completeness metric.

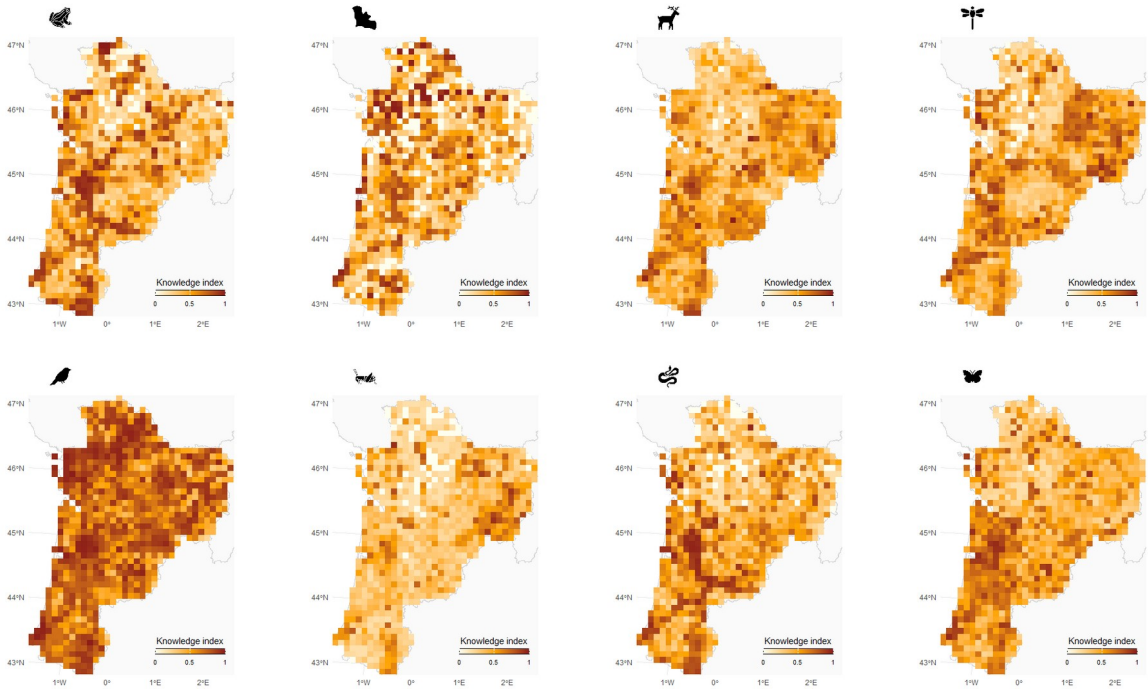

**Fig. S2** Distribution of knowledge gaps for each taxonomic group using the knowledge metric.